\documentstyle[12pt]{article}

\begin{document}
\baselineskip = 24pt

\begin{titlepage}

%%-shift up and right to accommodate the standard ICTP blue margin-%%
\hoffset = .5truecm
\voffset = -2truecm

\centering

\null
\vskip -1truecm

\vskip 1truecm

{\normalsize \sf \bf International Atomic Energy Agency\\
and\\
United Nations Educational, Scientific and Cultural Organization\\}
\vskip 1truecm
{\huge \bf
INTERNATIONAL CENTRE\\
FOR\\
THEORETICAL PHYSICS\\}
\vskip 3truecm

%%-Title and abstract page-%%
{\LARGE \bf
Continuous Local Symmetry in Ising-type Models
\\}
\vskip 1truecm

{\large \bf
S. Randjbar-Daemi
\\}
\medskip
{\large      and\\}
\medskip
{\large \bf
J. Strathdee
\\}

\vskip 7truecm

{\bf International Centre for Theoretical Physics \\}
January 1994
\end{titlepage}

%%-move to normal A4-%%
\hoffset = -1truecm
\voffset = -2truecm

\title{\bf
Continuous Local Symmetry in Ising-type Models
}

\author{
{\bf
S. Randjbar-Daemi
}\\
\normalsize International Centre for Theoretical Physics, Trieste 34100,
{\bf Italy}\\
{\normalsize and}\\
\normalsize
{\bf}\\
{\bf J. Strathdee}\\
\normalsize International Centre for Theoretical Physics, Trieste 34100,
{\bf Italy} }

\date{January 1994}
\newpage

\maketitle

\begin{abstract}
A class of generalized Ising models is examined with a view to extracting a low
energy sector
 comprising
 Dirac fermions coupled to Yang-Mills vectors. The main feature of this
approach is a set of
gap equations, covariant with respect to one of the $4$-dimensional
crystallographic space groups.

\end{abstract}

\newpage

%%-main body of paper-%%

Many authors have suggested that gauge fields may not be elementary
\cite{kn:Landau.}. In this note we would like
to reconsider this possibility in the context of a discrete Ising-type model.
We
argue that it is feasible to set up a self-consistent scheme from which both
gauge vectors and
 Dirac fermions emerge as effective dynamical variables in the continuum limit.

Models of the Ising type would seem to provide the most primitive kind of
structure
on which one could hope to develop a semi-realistic low energy physics. We
shall therefore consider
an array of sites, $A$, on each of which there is an Ising variable $S(A)=0,1$.
The partition
function is given by the sum over
configurations
%%%%%%%%%%%%%%%%%%%%%%%%%%%%%%%%%%%%%%%%%%%%%%%%%%
\begin{equation}
Z =\sum_{S}\ exp\left( -{\beta\over{2}}\sum_{A,B}\ S(A)J(A,B)S(B) +i\pi\
\sum_{A}\ S(A)\right)
\label{eq:Z}
\end{equation}
%%%%%%%%%%%%%%%%%%%%%%%%%%%%%%%%%%%%%%%%%%%%%
where the couplings, $J(A,B)$ are to be specified. We have in mind a
$4$-dimensional
 lattice in which the couplings are not necessarily restricted to nearest
neighbours. To restrict
 their form we shall assume that these couplings are invariant with respect to
one of the $4$-dimensional
crystallographic space groups. In addition, we shall suppose
that the sites are grouped into $N$- component cells such that the couplings
are sensitive
only to the cell and do not distinguish among the individual sites in a cell.
This
introduces a local permutation symmetry, $S_{N}$, into the model. With these
assumptions, the
partition function (\ref{eq:Z}) can be represented in the weak coupling, or
high temprature, limit
 in the form of an integral over Grassmann
 variables, $\psi_{i}(n)$, $\bar\psi_{i}(n)$, $i= 1,...,N$, $n\in Z^{4}$,
%%%%%%%%%%%%%%%%%%%%%%%%%%%%%%%%%%%%%%%%%%%%%%%%%%
\begin{equation}
Z =\int\ \left( d\psi d\bar\psi\right)\ exp\left(
-{1\over{2}}\sum_{n,m}\bar\psi(n)\psi(n) J(n-m)\bar\psi(m)\psi(m)\right)
\label{eq:PI.2}
\end{equation}
%%%%%%%%%%%%%%%%%%%%%%%%%%%%%%%%%%%%%%%%%%%%%
where the coupling function, $J(n)$, is invariant under the action of the
crystal point
group, $n\rightarrow gn$. Equation (\ref{eq:PI.2}) is the infinite temperature
limit of a fermionic
partition function. The external magnetic field ${\displaystyle
i\pi\over\displaystyle\beta}$ has been introduced in (\ref{eq:Z})
in order to render the boundary conditions of the finite temperature fermionic
path integral
periodic. In the Grassmann version (\ref{eq:PI.2}) the local permutation
 symmetry is enlarged to local $GL(N,C)$.

Our purpose is to study the low energy properties of the system
(\ref{eq:PI.2}). In
particular, we want to know if there is a non-vanishing condensate
$<\psi(n)\bar\psi(m)>$
and if so, what are its symmetries. In order to obtain fermionic excitations
carrying spin $1/2$ it is essential that the ground state symmetry should
involve
an embedding of the point group of the lattice into the local group, $GL(N,C)$,
such that
the Grassmann fields transform as spinors. It is for this reason that we assume
the point
group to belong to one of the crystal classes, i.e. be a discrete subgroup of
$O(4)$.

The largest $4$-dimensional crystallographic  point group, $H$, has $1152$
elements ( including
improper elements) \cite{kn:Hurley} \footnote{The corresponding crystal lattice
turns out to be
the root lattice of the exceptional algebra $F_{4}$ \cite{kn:Itzykson}.}.
Its spinor covering possesses a $4$-dimensional spinor among its
irreducible representations. We shall therefore assume that $N$ is a multiple
of
$4$ and that the ground state symmetry is $H\times U(N/4)$ together with the
lattice translations.
Under this group the $N$-component $\psi$ decomposes into "Dirac" spinors of
$H$, in the
fundamental representation of $U(N/4)$.

To study the condensation problem we formulate self-consistent, Hartree-type
equations for
the condensate amplitude
%%%%%%%%%%%%%%%%%%%%%%%%%%%%%%%%%%%%%%%%%%%%%%%%%%
\begin{equation}
c(n-m)=<\psi(n)\bar\psi(m)>
\label{eq:con.1}
\end{equation}
%%%%%%%%%%%%%%%%%%%%%%%%%%%%%%%%%%%%%%%%%%%%%
and require the solution to be invariant with respect to $H\times U(N/4)$. This
means, in
 particular the functions $c(s)$ should satisfy
%%%%%%%%%%%%%%%%%%%%%%%%%%%%%%%%%%%%%%%%%%%%%%%%%%
\begin{equation}
c(gs)=a(g)c(s)a(g^{-1})
\label{eq:con.2}
\end{equation}
%%%%%%%%%%%%%%%%%%%%%%%%%%%%%%%%%%%%%%%%%%%%%
where $g$ is an element of the point group and $a(g)$ is its image in the local
group, $GL(N,C)$.

A convenient way to obtain the self-consistency conditions is by introducing a
bosonic
auxiliary field, $\chi(n,m)$, and replacing (\ref{eq:PI.2}) by the equivalent
form,
%%%%%%%%%%%%%%%%%%%%%%%%%%%%%%%%%%%%%%%%%%%%%%%%%%
\begin{equation}
\begin{array}{lll}
Z &=\int\ \left( d\psi d\bar\psi\right)(d\chi)\ exp\left[-\sum_{n,m}
\left(
\psi(n)\chi(n,m)\bar\psi(m) +
{\displaystyle tr\left(\chi(n,m)\chi(m,n)\right)\over\displaystyle 2J(n-m)}
\right)\right]\\
&\phantom{aa}\\
&=\int(d\chi)\ exp\ \left[- ln\ Det\ \chi\ -\ \sum_{n,m}
{\displaystyle tr\left(\chi(n,m)\chi(m,n)\right)\over\displaystyle
2J(n-m)}\right]
\end{array}
 \label{eq:Hubb.1}
\end{equation}
%%%%%%%%%%%%%%%%%%%%%%%%%%%%%%%%%%%%%%%%%%%%%
where the latter version is obtained by integrating the Grassmann variables. In
the Hartree
approximation, the condensate amplitude, $c(n-m)= <\chi(n,m)>$, is given by the
saddle points
of (\ref{eq:Hubb.1}). These points are obtained by solving the "gap equations",
%%%%%%%%%%%%%%%%%%%%%%%%%%%%%%%%%%%%%%%%%%%%%%%%%%
\begin{equation}
c(s)=J(s)G(s)
 \label{eq:vev.1}
\end{equation}
%%%%%%%%%%%%%%%%%%%%%%%%%%%%%%%%%%%%%%%%%%%%%
where $G$ is the fermion propagator. This propagator is to be expressed in
terms of $c(s)$,
%%%%%%%%%%%%%%%%%%%%%%%%%%%%%%%%%%%%%%%%%%%%%%%%%%
\begin{equation}
G(n-m)\ =\ \int\ {\displaystyle d^{4}k\over\displaystyle (2\pi)^{4}}\ \tilde
G(k)\ e^{ik(n-m)}
 \label{eq:fun.}
\end{equation}
%%%%%%%%%%%%%%%%%%%%%%%%%%%%%%%%%%%%%%%%%%%%%
where the integral extends over a cell of volume $(2\pi)^{4}$, a Brillouin
zone. The inverse of
$\tilde G$ is given by
%%%%%%%%%%%%%%%%%%%%%%%%%%%%%%%%%%%%%%%%%%%%%%%%%%
\begin{equation}
\tilde G^{-1}(k)=\sum_{s} c(s)e^{-iks}
 \label{eq:fun.x}
\end{equation}
%%%%%%%%%%%%%%%%%%%%%%%%%%%%%%%%%%%%%
It has the same invariance properties as $c(s)$ since the lattice sum in
(\ref{eq:fun.x}) is
invariant under the action of $H$. Under this action the lattice decomposes
into
non-intersecting orbits. On each orbit $J(s)$ is a constant and $c(s)$ is
expressible in terms
of a finite number of parameters. These parameters are determined as functions
of the
couplings by solving the gap equations (\ref{eq:vev.1}) or, equivalently, by
searching for
extrema of the energy density
$$
V\ =\ \sum_{s}\ {\displaystyle tr\left(c(s)c(-s)\right)\over\displaystyle
2J(s)}\ -\
\int\ {\displaystyle d^{4}k\over\displaystyle (2\pi)^{4}}\ ln\ det\  \tilde
G(k)^{-1}
$$

We have verified that for the nearby orbits, $s^{2}=2, 4$ and $6$, the
invariance condition
(\ref{eq:con.2}) implies
%%%%%%%%%%%%%%%%%%%%%%%%%%%%%%%%%%%%%%%%%%%%%%%%%%
\begin{equation}
 c(s)=c_{1}(s)+c_{2}(s)s_{\mu}\gamma_{\mu}
\label{eq:g.s. sol.1}
\end{equation}
%%%%%%%%%%%%%%%%%%%%%%%%%%%%%%%%%%%%%%%%%%%%%
where $c_{1}$ and $c_{2}$ are orbit invariants and $\gamma_{\mu}$ are Dirac
matrices. For more
distant orbits there would be some proliferation of parameters. However,
insofar as only the
nearby orbits are involved, the general form of the
fermion propagator is given by
%%%%%%%%%%%%%%%%%%%%%%%%%%%%%%%%%%%%%%%%%%%%%%%%%%
\begin{equation}
\begin{array}{llll}
 \tilde G^{-1}(k)\ =\ M(k)+i\gamma_{\mu}Q_{\mu}
\end{array}
\label{eq:Fer.prop.1}
\end{equation}
%%%%%%%%%%%%%%%%%%%%%%%%%%%%%%%%%%%%%%%%%%%%%
where $M$ and $Q_{\mu}$ represent the invariant sums,
%%%%%%%%%%%%%%%%%%%%%%%%%%%%%%%%%%%%%%%%%%%%%%%%%%
\begin{equation}
\begin{array}{llll}
M(k)=\sum_{s}c_{1}(s)\ e^{-iks}\ \ \ \ \ ,\ \ \ \ \
Q_{\mu}(k)= {\partial\over\partial k_{\mu}}\sum_{s}c_{2}(s)\ e^{-iks}
\end{array}
\label{eq:Fer.prop.2}
\end{equation}
%%%%%%%%%%%%%%%%%%%%%%%%%%%%%%%%%%%%%%%%%%%%%

It is known \cite{kn:Itzykson} that the group $H$ allows four algebraically
independent invariants to be constructed from
a single $4$-vector, $k_{\mu}$. They are of order $k^{2}$, $k^{6}$, $k^{8}$ and
$k^{12}$.
Since $k^{2}$
is also $O(4)$ invariant, it follows that $\tilde G^{-1}(k)$ is $O(4)$
invariant in the neighbourhood
of $k_{\mu}=0$ if terms of order $k^{6}$ are neglected, i.e.
%%%%%%%%%%%%%%%%%%%%%%%%%%%%%%%%%%%%%%%%%%%%%%%%%%
\begin{equation}
\begin{array}{llll}
 \tilde G^{-1}(k)=Z_{2}^{-1}\left(m+i\not k\right)
\end{array}
\label{eq:Fer.prop.3}
\end{equation}
%%%%%%%%%%%%%%%%%%%%%%%%%%%%%%%%%%%%%%%%%%%%%
where $Z_{2}^{-1}$ and $Z_{2}^{-1} m$ are linear in the condensate parameters
$c_{1}$ and
$c_{2}$. When the gap equations are solved they will be expressed in terms of
the $J$' s. One
can hope that the fermion mass, $m$, can be made small by tuning these
couplings.

The condensation phenomenon, which yielded propagating Dirac fermions, will
also give rise to propagating
bosons. These bosons are associated with the fluctuations of the "auxiliary"
field, $\chi(n,m)$.
In particular,  because of the local symmetry
one expects to find Yang-Mills vectors by means of the ansatz,
%%%%%%%%%%%%%%%%%%%%%%%%%%%%%%%%%%%%%%%%%%%%%%%%%%
\begin{equation}
\begin{array}{llll}
 \chi(x, x-s)\ =\ c(s)\ T\ \left(\ exp\ i\ \int_{x-s}^{x}\ A\right)
\end{array}
\label{eq:W.an}
\end{equation}
%%%%%%%%%%%%%%%%%%%%%%%%%%%%%%%%%%%%%%%%%%%%%
which should be valid in the continuum limit. In this limit, the vector
$A_{\mu}$ transforms
in the usual way,
%%%%%%%%%%%%%%%%%%%%%%%%%%%%%%%%%%%%%%%%%%%%%%%%%%
\begin{equation}
\begin{array}{llll}
 \delta\ A_{\mu}(x)\ =\ \partial_{\mu}\theta(x)\ -i\left[\
A_{\mu}(x),\theta(x)\ \right]
\end{array}
\label{eq:Y.M. trans.0}
\end{equation}
%%%%%%%%%%%%%%%%%%%%%%%%%%%%%%%%%%%%%%%%%%%%%
where $\theta(x)$ belongs to the algebra of $U(N/4)$, commutes with $c(s)$, and
is slowly varying
on the lattice scale, $|\partial_{\mu}\theta |<<|\theta |$. On substituting the
ansatz
(\ref{eq:W.an}) into the bosonic action (\ref{eq:Hubb.1}), expanding in powers
of $A$,
$\partial A$,..., the leading terms must compose themselves into the standard
Yang-Mills form. To
identify the gauge coupling constant it is sufficient to examine the bilinear
term
%%%%%%%%%%%%%%%%%%%%%%%%%%%%%%%%%%%%%%%%%%%%%%%%%%
\begin{equation}
{1\over 2}\int\ {\displaystyle d^{4}k\over\displaystyle (2\pi)^{4}}\
\sum_{\alpha}\ \tilde A^{\alpha}_{\mu}(-k)
 \tilde\Delta^{-1}_{\mu \nu}(k)\tilde A^{\alpha}_{\nu}(k)
\label{eq:bilin. Y.M}
\end{equation}
%%%%%%%%%%%%%%%%%%%%%%%%%%%%%%%%%%%%%%%%%%%%%
where the sum over  $\alpha=1,2,..,(N/4)^{2}$ accounts for the local symmetry
$U(N/4)$. Invariance
with respect to the gauge transformations (\ref{eq:Y.M. trans.0}) implies
%%%%%%%%%%%%%%%%%%%%%%%%%%%%%%%%%%%%%%%%%%%%%%%%%%
\begin{equation}
\tilde\Delta^{-1}_{\mu\nu}={1\over g^{2}}\left(k^{2}\delta_{\mu\nu}
-k_{\mu}k_{\nu}\right)+...
\label{eq: Y.M.c.c}
\end{equation}
%%%%%%%%%%%%%%%%%%%%%%%%%%%%%%%%%%%%%%%%%%%%%
near $k=0$. To express the coupling constant , $g^{2}$, in terms of the
condensate parameters,
$c_{1}$, $c_{2}$, it would be necessary to expand the coefficients,
%%%%%%%%%%%%%%%%%%%%%%%%%%%%%%%%%%%%%%%%%%%%%%%%%%
\begin{equation}
\begin{array}{llll}
&\tilde\Delta^{-1}_{\mu\nu}(k)=\sum_{s,s^{\prime}}\
{\displaystyle {1-e^{iks}}\over\displaystyle ks}\ s_{\mu}\\
&\phantom{aa}\\
&\left[
{\displaystyle \delta_{s+s^{\prime},0}\over\displaystyle J(s)}
tr
\left(
c(s)c(-s)
\right)
+
\int\ {\displaystyle d^{4}q\over\displaystyle (2\pi)^{4}}\
tr
\left(
c(s)\tilde G(q)c(s^{\prime})\tilde G(q-k)
\right)
e^{-iq(s+s^{\prime})}
\right]
{\displaystyle {1-e^{iks^{\prime}}}\over\displaystyle
ks^{\prime}}s^{\prime}_{\nu}
\end{array}
\label{eq:inv. Y. M.prop.}
\end{equation}
%%%%%%%%%%%%%%%%%%%%%%%%%%%%%%%%%%%%%%%%%%%%%
around $k=0$.

The ansatz (\ref{eq:W.an}) can also be substituted into the fermionic part of
the action defined by
(\ref{eq:Hubb.1}) to obtain the coupling of the gauge field to the fermions.
One finds
the usual minimal coupling. To summarize, the low energy sector of the
$4$-dimensional
generalized Ising
model is describe by the effective action
%%%%%%%%%%%%%%%%%%%%%%%%%%%%
\begin{equation}
 S=\int d^{4}x
\left[
{{1}\over {4g^{2}}}F^{\alpha}_{\mu\nu}F^{\alpha}_{\mu\nu} +
Z_{2}^{-1}\overline\psi\left(
\not\nabla +m\right)\psi\right]
\label{eq: Standard1}
\end{equation}
%%%%%%%%%%%%%%%%%%%%%%%%%%%%%%%%%%%%%%%%
where $g^{2}$ , $Z_{2}^{-1}$ and $m$ are ultimately to be expressed in terms of
the
lattice couplings, $J(s)$, by solving the gap equations (\ref{eq:vev.1}) and
substituting
the resulting condensate parameters into (\ref{eq:Fer.prop.2})  and
(\ref{eq:inv. Y. M.prop.}).

In this note we have attempted to minimize the number of assumptions about the
Ising model coupling
constants consistent  with obtaining a low energy structure like (\ref{eq:
Standard1}). In particular, we have
not restricted the $J$' s to the nearest neighbours or next nearest , etc. Our
reason for doing this is the
hope that it will eventually be possible to treat the $J$' s as dynamical
variables : quantities to be
obtained by solving equations of motion. The $4$-dimensional lattice structure
should then
emerge as a feature of the ground state symmetry. The space group would then
have a
status analogous to that of the Poincare group in general relativity. Indeed,
such a theory
would have to yield a long range tensorial interaction somewhat like
gravitation.

\end{document}